\documentstyle[aps,pre,manuscript]{revtex}

\begin{document}
\draft
\title{Response of an electron system to a periodic potential}
\author{S. Nagy, K. Sailer}
\address{ Department of Theoretical Physics, Debrecen University,
        Debrecen, Hungary }
\date{\today}
\maketitle

\begin{abstract}
We give a quantum field theoretical treatment of a one
dimensional electron system with a fixed chemical potential $\mu$.
The non-perturbative Lindhard response function is found for an
electron system in a sinusoidal potential.
\end{abstract}


\section{Introduction}
 Our goal is to treat relativistically the response of an electron 
system to a  periodic electric field in the framework
of 1+1 dimensional QED (QED$_2$).
We assume the
presence of a classical pe\-ri\-o\-dic electric mean field, 
i.e. the electromagnetic
field variable $A_\nu(x)=\bar A_\nu+\alpha_\nu$ is the sum of this 
classical field and the quantum fluctuations.
The interaction of the fermions with this classical field is treated exactly,
whereas the quantum fluctuations
of the electromagnetic field are taken into account at 1-loop order.
Requiring that the classical field is a mean field enables us to 
investigate a Wigner crystal
(Wigner 1934, Grimes 1979) relativistically.
To get a better understanding of such an electron system, it is
unavoidable to investigate its non-perturbative response to a classical
periodic electric field, that is the issue of the present work.

The classical periodic electric field is chosen in the form:
\begin{equation}
{\bar A}^\nu (x) = a \ \delta^\nu_0\cos(\ell x)+\delta^\nu_0\mu,
\label{meanfield}
\end{equation}
with amplitude $a$ and wavenumber $\ell$.
The homogeneous term $\mu$ corresponds to the chemical potential of the
system.

\section{Method}

In QED the expectation value of the gauge field vanishes.
To take into account a mean field self-consistently in this theory
we should modify its generating functional.

Our method is based on defining a sector of QED for field configurations
belonging to quantum fluctuations around a given vacuum expectation
value $\langle A^\mu (x) \rangle$.
We multipy the integrand of the common generating functional of
QED$_2$ by the factor 1 written in the form
\begin{eqnarray}
 1 &=& \int dc \ \delta ( \int dx A_\mu n^\mu - c \Omega),
\end{eqnarray}
where $n^\mu (x)$ is a vector in the space of the vector potential
configurations and with $\Omega = TV$ the spacetime volume. Then, we find
\begin{eqnarray}
  Z_{QED} &=& \int dc Z_{QED} ' \lbrack n,c \rbrack ,
\end{eqnarray}
where the functional
\begin{eqnarray}
   Z_{QED} ' \lbrack n,c \rbrack &=&
    \int {\cal D} {\bar \psi} {\cal D} \psi {\cal D} A 
   \exp \left\{ {\rm i} S_{em} \lbrack A, \xi \rbrack
      +  {\rm i} S_D \lbrack A, {\bar \psi}, \psi \rbrack 
        \right\}
    \delta  ( \int dx A_\mu n^\mu - c \Omega)
\end{eqnarray}
is the generating functional of the sector
belonging to vector potential configurations in a hypersurface
orthogonal to $n^\mu (x)$. $A^\mu (x)$ is the electromagnetic 
and $\psi (x)$ is the Dirac field,
$S_{em} \lbrack A, \xi \rbrack $
and $ S_D \lbrack A, {\bar \psi}, \psi \rbrack$ are the action of the
electromagnetic field and the Dirac action, resp., and $\xi$ is the
gauge fixing parameter.
By shifting the integration variable according to $\alpha^\mu (x) =
A^\mu (x) - {\bar A}^\mu (x) $, the projected generating functional 
becomes
\begin{eqnarray}
\label{zqedp}
    Z_{QED} ' \lbrack {\bar A},c \rbrack &=&
  \int d\lambda \int {\cal D} \alpha 
  \exp \left\{ {\rm i} S_{em} \lbrack {\bar A} + \alpha , \xi \rbrack
       \right\}
    \exp \left\{  
   {\rm i} \lambda \left( \int dx {\bar A}^\mu {\bar A}_\mu + \int dx {\bar
     A}^\mu \alpha_\mu - c \Omega \right)
           \right\}\times
      \nonumber\\
  &   &
   \int {\cal D} {\bar \psi} {\cal D} \psi
    \exp \left\{ {\rm i} S_D \lbrack {\bar A} + \alpha , {\bar \psi}
      , \psi \rbrack  \right\} .
\end{eqnarray}
Self-consistency requires that
\begin{eqnarray}
\label{constraint}
  \langle A^\mu (x) \rangle = {\bar A}^\mu (x)
        , \qquad {\mbox {i.e.}} \qquad
    \langle \alpha^\mu \rangle =0 .
\end{eqnarray}
This condition fixes the value of $c=\int dx \bar A_\mu\bar A^\mu$.
To fulfill the requirement of (\ref{constraint}) an extra 
renormalization condition is needed to every single loop correction.
In our treatment the form of $\bar A_\nu$ coincides with (\ref{meanfield}).
Calculations of any physical quantity is performed by taking an expectation
value of that quantity with the modified generating functional.

To determine the Lindhard function we should calculate the 
polarization $\Pi (x,x')$ the form of which in QED is:
\begin{equation}
\Pi_{\mu\nu}(x,x') = \frac{e^2}{i}\mbox{tr}
                (\gamma_\mu G(x,x') \gamma_\nu G(x',x)),
\end{equation}
where $G(x,x')$ is the electron propagator.

Diagrammatically $\Pi (x,x')$ corresponds to the diagram in FIG. 1.
(without the external legs).
In FIG. 1. the solid lines represent the electron 
propagator and the wavy lines 
correspond to the photon propagator.
Since we are interested first in the response of the electron system
to an arbitrary periodic potential, $a$ and $\ell$ should be
considered as free parameters in the Lindhard function.

\section{RESULTS}

The electron propagator
belonging to the projected generating functional contains all
the effects of the periodic classical field and the  propagator
itself looses its translational symmetry due to this sinusoidal
vector potential. Another consequence of the periodic classical field
is that the electron propagator from (\ref{zqedp})
does not have analytic form, the
Linhard response function can only be obtained numerically.
First, we write down $\Pi_{\mu\nu}$ in momentum space according to:
\begin{eqnarray}
\Pi_{\mu\nu}(k,k') &=& \frac{e^2}{i}\int dx\int dx' \ 
                e^{ikx+ik'x'}\mbox{tr}
                (\gamma_\mu G(x,x') \gamma_\nu G(x',x)).
\label{Pikkp}
\end{eqnarray}
In momentum space the two-vector $k=(k_o,\vec k)$ is the momentum 
of the photon in FIG. 1. coming from the left, while $k'=(k'_o,\vec k')$
denotes the momentum of the outgoing photon.
Performing the integrations in (\ref{Pikkp}) we get:
\begin{eqnarray}
\Pi_{\mu\nu}(\vec k,k_0;\vec k',k'_0{=}-k_0) &=&
        -{e^2}\sum_{k_1s_1k_2s_2}\sum_{n_1n_2n_3n_4}
        \frac1{\epsilon_{k_1s_1}^{(-)}+\epsilon_{k_2s_2}^{(+)}-2k_0}
        \times\nonumber\\
        &&\delta(\vec k+\vec k'+\vec \ell
                (n_3+n_4-n_1-n_2))\times
        \nonumber\\
&&
        \Bigl({\cal J}_{\mu\nu} \delta(\vec k_1+\vec k_2-\vec k+
                \vec \ell(n_1+n_2))+
        \nonumber\\
&&
        {\cal J}_{\nu\mu} \delta(\vec k_1+\vec k_2-\vec k'+
                \vec \ell(n_1+n_2))\Bigr),
\label{Pi}
\end{eqnarray}
where $\epsilon_{k_1s_1}^{(-)}$ and $\epsilon_{k_2s_2}^{(+)}$ are 
positive energy eigenvalues corresponding to
the positive and negative frequency eigenspinors of the Dirac Hamiltonian
$H_D[\bar A]$
with the potential (\ref{meanfield}), and ${\cal J_{\mu\nu}}$ contains
products of the eigenspinors of $H_D[\bar A]$ in the following form:
\begin{eqnarray}
{\cal J_{\mu\nu}} &=& \bar v^{k_2s_2}_{n_2} \gamma_\mu u^{k_1s_1}_{n_1}
                \cdot \bar u^{k_1s_1}_{n_3} \gamma_\nu u^{k_2s_2}_{n_4}.
\end{eqnarray}
$u^{ks}_{n}$ and $v^{ks}_{n}$ are parts of the Bloch type eigenspinors
(Nagy 2000).
The first Dirac delta in (\ref{Pi}) expresses the breaking of the
translational symmetry due to the periodic field.
We expect that (\ref{Pi}) at close to $k_0=k'_0=0$ has a maximum value for
$\vec k = \vec \ell$ when the Fermi surface is positioned close
to the boundary of the Brillouin zone.
It would mean that there is no energy exchange
between the two electrons in FIG. 1. but they exchange momentum 
$\vec\ell$. It reminds us to the
so called Peierls phase transition where near the Fermi surface electrons
can jump from one side of the Fermi surface to the other without any energy
cost. The instability caused by this procedure is stabilized by a gap in
the dispersion relation arising at the edge of the Fermi surface.
In our treatment the stability is ensured by the periodic electric 
mean field which produces a gap at the edge of the Brillouin zone.

\section*{ACKNOWLEDGEMENTS}

The authors would like to thank J. Polonyi and
G. Plunien for the useful discussions.
This work was supported by the projects OTKA T023844/97, DAAD-M\"OB 27/1999
and NATO SA( PST.CLG 975722)5066.
Numerical computations are in progress on the computer cluster donated by
the Alexander von Humboldt Foundation.

\section*{REFENCES}
\noindent
Grimes, C. C. and Adams, G., Phys. Rev. Lett. {\bf 42}, 795 (1979)\\
Nagy, S., Sailer, K., Heavy Ion Phys. {\bf 11}, Nos 1-2, 67 (2000)\\
Wigner, E. P., Phys. Rev. {\bf 46}, 1002 (1934)

\section*{FIGURE CAPTIONS}

FIG. 1.: Diagram of $\Pi_{\mu\nu}(x,x')$.

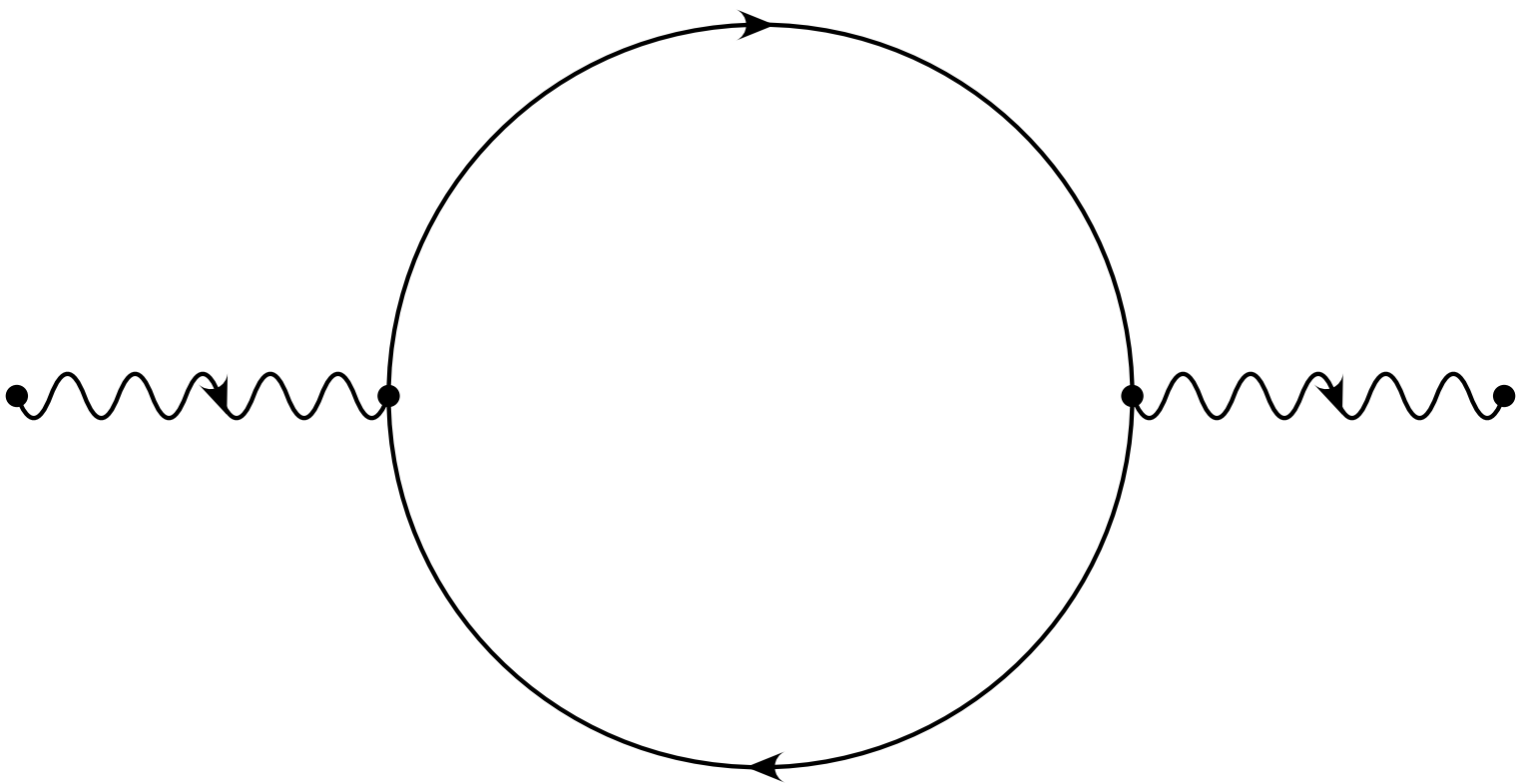
\end{document}